\documentclass[twocolumn,prb,superscriptaddress,amsmath,amssymb,aps]{revtex4}

\usepackage[utf8]{inputenc}

\usepackage{dcolumn}
\usepackage{bm}
\usepackage{amsmath}
\usepackage{amssymb}
\usepackage{subfigure} 
\usepackage{tikz}
\usepackage{verbatim}

\usetikzlibrary{arrows,shapes}
\tikzstyle{block}     = [rectangle, draw, text centered, rounded corners, minimum height=2em, minimum width=2em]
\tikzstyle{connector} = [->,thick]
\tikzstyle{line}      = [draw, -latex']
\tikzstyle{cloud}     = [draw, ellipse, node distance=3cm, minimum height=2em]
\tikzstyle{blank}     = [rectangle, height=0.1em, width=0.1em]

\newcommand{\braopket}[3]{{\langle #1|#2|#3\rangle}}

\DeclareGraphicsExtensions{.pdf,.png,.eps}

\begin{document}

\title{BOPcat software package for the construction and testing of tight-binding models and bond-order potentials}

\author{A.N.~Ladines}
\affiliation{Atomistic Modelling and Simulation, ICAMS, Ruhr-Universit\"at Bochum, D-44801 Bochum, Germany}

\author{T.~Hammerschmidt}
\affiliation{Atomistic Modelling and Simulation, ICAMS, Ruhr-Universit\"at Bochum, D-44801 Bochum, Germany}

\author{R.~Drautz}
\affiliation{Atomistic Modelling and Simulation, ICAMS, Ruhr-Universit\"at Bochum, D-44801 Bochum, Germany}

\begin{abstract}
Atomistic models like tight-binding (TB), bond-order potentials (BOP) and classical potentials describe the 
interatomic interaction in terms of mathematical functions with parameters that need to be adjusted for a particular material. 
The procedures for constructing TB/BOP models differ from the ones for classical potentials. We developed the BOPcat software 
package as a modular python code for the construction and testing of TB/BOP parameterizations. It makes use of atomic 
energies, forces and stresses obtained by TB/BOP calculations with the BOPfox software package. It provides a graphical 
user interface and flexible control of raw reference data, of derived reference data like defect energies, of automated 
construction and testing protocols, and of parallel execution in queuing systems. We outline the concepts and usage of the 
BOPcat software and illustrate its key capabilities by exemplary constructing and testing of an analytic BOP for Fe. The 
parameterization protocol with a successively increasing set of reference data leads to a magnetic BOP that is transferable 
to a variety of properties of the ferromagnetic bcc groundstate and to crystal structures that were not part of the training set.
\end{abstract}

\maketitle

\section{Introduction}

Atomistic simulations are indispensable in materials science and require a robust description of the interatomic interaction. 
The application of highly accurate density functional theory (DFT) calculations is often limited by the computationally involved description of the electronic structure. 
A systematic coarse-graining of the electronic structure~\cite{Drautz2006,Drautz2011} leads to the tight-binding (TB) bond model~\cite{Sutton-88} and to analytic bond-order potentials (BOP)~\cite{Hammerschmidt-09-IJMR,Drautz2015}.
These TB/BOP models can be evaluated at significantly reduced numerical effort and provide a transparent and intuitive model of the interatomic bond for the prediction of materials properties, see e.g. Refs.~\cite{Mrovec2004,Mrovec2007,Mrovec2011,Seiser-11-2,Cak2014,Ford-14,Ford2015,Wang-19}. 

The TB/BOP models employ adjustable parameters that need to be optimized for a particular material.
This optimization is in principle comparable to the parameterization of classical potentials which can be performed with various existing software packages~\cite{Brommer2007,Duff2015,Barrett2016,Stukowski2017}.
The parameterization of TB/BOP models, however, requires sophisticated successive optimization steps, computationally-efficient handling of large data sets and an interface to a TB/BOP calculator~\cite{Horsfield-96,Hammerschmidt2019}.

Here, we present BOPcat (\textbf{B}ond-\textbf{O}rder \textbf{P}otential \textbf{c}onstruction \textbf{a}nd \textbf{t}esting), a software to parameterize TB/BOP models as implemented in the BOPfox software~\cite{Hammerschmidt2019}. 
The parameters of the models are optimized to reproduce target properties like energies, forces, stresses, eigenvalues, defect formation energies, elastic constants, etc. 
With the interface of BOPfox to LAMMPS~\cite{Plimpton-95} and ASE~\cite{Larsen2017}, the list of target properties can in principle be extended to include dynamical properties. 
We illustrate the capability of the BOPcat software by constructing and testing an analytic BOP with collinear magnetism for Fe. 
Extensive tests show the good transferability of the BOP to properties which were not included in the parameterization, particularly to elastic constants, point defects, $\gamma$ surfaces, phonon spectra and deformation paths of the ferromagnetic bcc groundstate and to other crystal structures.
The structures and target properties used here are taken from DFT calculations but could also include experimental data or other data sources. 

We first provide a brief introduction of bond-order potentials in Sec.~\ref{sec:BOPformalism}. 
In Sec.~\ref{sec:programflow}, the BOPcat program is outlined and the implemented modules are described.
In Sec.~\ref{sec:application}, we discuss the construction of an analytic BOP for Fe and its testing. In the appendix we provide
further examples of parameterization protocols for BOPcat.

\section{Bond-order potential formalism}\label{sec:BOPformalism}

Analytic BOPs provide a robust and transparent description of the interatomic interaction that is based on a coarse-graining of the electronic structure~\cite{Drautz2006,Drautz2011,Drautz2015} from DFT to TB to BOP.
In the following we compile the central equations and functions of the TB/BOP formalism that are parameterized in BOPcat. 

In the TB/BOP models, the total binding energy is given by
\begin{equation}
 \label{eq:U_B_general}
  U_B  = U_{\rm{bond}} + U_{\rm{prom}} + U_{\rm{ion}} + U_{\rm{es}} + U_{\rm{rep}} + U_{\rm{X}}
\end{equation}
with the bond energy $U_{\rm{bond}}$ due to the formation of chemical bonds, the promotion energy $U_{\rm{prom}}$ from redistribution of electrons across orbitals upon bond formation, the onsite ionic energy $ U_{\rm{ion}}$ to charge an atom, the intersite electrostatic energy $U_{\rm{es}}$, 
the exchange energy $U_{\rm{X}}$ due to magnetism and the repulsive energy $U_{\rm{rep}}$ that includes all further terms of the second-order expansion of DFT. 
The individual energy and force contributions are described in detail in Ref.~\cite{Hammerschmidt2019}, their functional forms and according parameters for the exemplary construction of a magnetic BOP for Fe are given in Sec.~\ref{sec:application}.

The bond energy is obtained by integration of the electronic density of states (DOS)  $n_{i\alpha}$,
\begin{equation}
 U_\mathrm{bond} = 2\sum_{i\alpha}\int^{E_F} (E-E_{i\alpha})n_{i\alpha}(E)dE
\end{equation}
with atomic onsite levels $E_{i\alpha}$ for orbital $\alpha$ of atom $i$.
In TB calculations, the DOS is obtained by diagonalization of the Hamiltonian $\hat{H}$ for a given structure.
In analytic BOPs~\cite{Drautz2006,Drautz2011,Drautz2015}, the DOS is determined analytically from the moments 
\begin{eqnarray}
 \label{eq:mu}
 \mu_{i\alpha}^{(n)} &=& \braopket{i\alpha}{\hat{H}^n}{i\alpha} \nonumber \\ 
                     &=&H_{i\alpha j\beta}H_{j\beta k\gamma}H_{k\gamma\ldots}\dotso H_{\ldots i\alpha} \nonumber  \\
                     &=&\int E^{n} n_{i\alpha}(E)dE
\end{eqnarray}
that are computed from the matrix elements $H_{i\alpha j\beta}$ of the Hamiltonian between pairs of atoms.
The BOP DOS provides an approximation to the TB DOS at a computational effort for energy and force calculations that scales linearly with the number of atoms~\cite{Teijeiro-16-2}.
The quality of the BOP approximation can be improved systematically by including higher moments with a power-law scaling for the increase in computational effort~\cite{Teijeiro-16-1}.
A detailed introduction to TB/BOP calculations in BOPfox is given in Ref.~\cite{Hammerschmidt2019}.

\section{Program Flow}\label{sec:programflow}

\begin{figure*}[t]
 \centering
 \includegraphics[width=1.9\columnwidth]{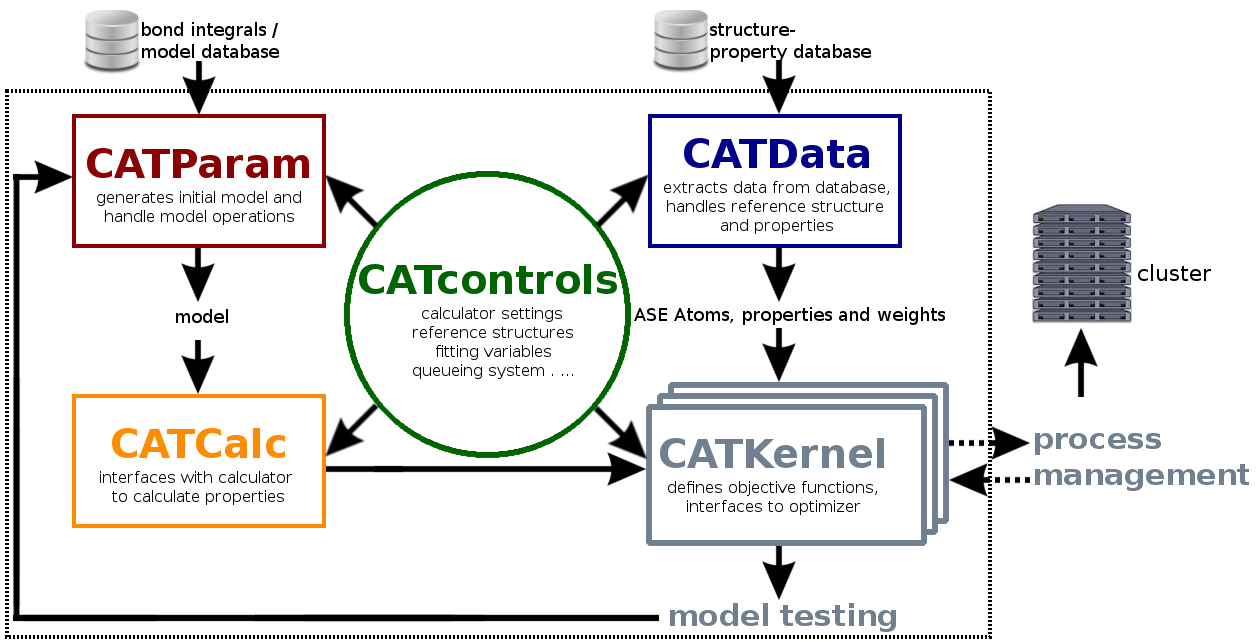}
 \caption{Workflow of the core optimization process in BOPcat. The input controls are specified and initialized.
          The reference data is read from a database and the reference and
          testing sets are generated. The initial model is read from a model database.
          The structures and initial model are assigned to the calculator interface.
          The calculator, reference properties and weights are then used to build the 
          optimization kernel. The kernel can be run as standalone process or be sent to 
          the process management of a queuing system.}
 \label{fig:workflow}         
\end{figure*}

BOPcat is a collection of Python modules and tools for the construction and testing of TB/BOP models.
The individual steps are specified by the user as a protocol in terms of a Python script. 
The overall construction and testing proceeds as follows:
\begin{itemize}
 \item define and initialize input controls (CATControls)
 \item generate initial model (CATParam)
 \item read reference data (CATData) 
 \item initialize calculator interface (CATCalc)
 \item build and run optimization kernel (CATKernel)
\end{itemize}
The individual tasks are modularized in the modules CATControls, CATParam, CATData, CATCalc and CATKernel that interact as illustrated in Fig.~\ref{fig:workflow}.
Due to its modular structure, one can also use only a subset of the modules, e.g., for successive optimizations.

In the following, the individual modules are described in detail with snippets of the python protocol for a basic construction process of a magnetic BOP for Fe.
Further example protocols are given in the appendix.

\subsection{Input controls: CATControls}
\label{sec:CATControls}

The variables required for running BOPcat are defined in the module CATControls. 
These include the list of chemical symbols for which the model is constructed, calculator settings, filters for the reference data, optimization variables, specifications of the model, etc.. 
The consistency of the input variables is checked within the module to catch inconsistent settings at an early stage. 
The CATControls object is then passed on to succeeding modules from which the relevant variables are assigned. 
As an example initialization of the CATControls object we define the functions of the TB/BOP model and the initial guess for the according parameters by providing an existing model from a file (\verb+models.bx+).

{\footnotesize
\begin{verbatim}
 from bopcat.catcontrols import CATcontrols
 ctr = CATControls()
 ctr.elements = ['Fe']
 ctr.model_pathtomodels = 'models.bx'
 ctr.calculator_nproc = 4
 ctr.calculator_parallel = 'multiprocessing'
 ctr.calculator_settings = {'scfsteps':500}
 ctr.data_filename = 'Fe.fit'
 ctr.data_free_atom_energies = {'Fe':-0.689}
 ctr.data_system_parameters = {'spin':[1,2],
                               'system_type':[0]}
 ctr.opt_optimizer = 'leastsq'
 ctr.opt_variables = [{'bond':['Fe','Fe'],
                       'rep1':[True,True]}]
 \end{verbatim}
}

Alternatively, BOPcat can generate an initial guess from raw bond integrals~\cite{Jenke2019} with additional user-specified information on the functions to be used for the individual energy contributions, the valence-type ($s$/$p$/$d$) and number of valence electrons of the elements, the cut-off radii for determining pairs of interacting atoms, etc. (see ~\ref{sec:app2}) 

\subsection{Reference data: CATData}
\label{sec:CATData}

The set of reference data of structures and properties is read from a text file by the CATData module.
The entries for reference data have the following format:

{\footnotesize
\begin{verbatim}
data_type = 0
code = 7
basis_set = 0
pseudopotential = 30
strucname = bcc
a1 = -1.3587  1.3587  1.3587
a2 =  1.3587 -1.3587  1.3587
a3 =  1.3587  1.3587 -1.3587
coord = cartesian
Fe    0.0000  0.0000  0.0000
energy = -7.9561
forces 0.0000 0.0000 0.0000
stress = -0.08423 -0.0843 -0.0843 -0. -0. -0.
system_type = 0
space_group = 229
calculation_type = 1
stoichiometry = Fe
calculation_order = 10.033
weight = 1.0
spin = 1
\end{verbatim}
}

In contrast to optimization procedures for classical potentials, the optimization of TB/BOP models with BOPcat can be carried out not only for atomic properties but also at the electronic-structure level by providing  eigenvalues and corresponding $k$-points as reference data. 
The optimization can be further extended to other properties of TB/BOP calculations (e.g. band width, magnetic moment) by supplying the according keyword and the data type of the BOPfox software in the reference data module of BOPcat.

CATData stores the reference data in a set of \verb+Atoms+ objects of the ASE python framework for atomistic simulations~\cite{Larsen2017}. 
Each of the ASE \verb+Atoms+ objects can have calculator-specific identifiers such as the settings of the DFT calculation (\verb+code+, \verb+basis_set+, \verb+pseudopotential+) as well as structure-related identifiers such as calculation type (\verb+calculation_type+) and calculation order (\verb+calculation_order+), which are used as filters to use only a subset of the available reference data.
A complete description of the implemented identifiers and filters is given in the manual of BOPcat. 
The corresponding reference data can include basic quantities like energies, forces, stresses as well as derived quantities such as defect energies. 
Extracting the reference structures and their corresponding properties is illustrated in the following: 

{\footnotesize
\begin{verbatim}
 from bopcat.catdata import CATData
 data = CATData(controls=controls)
 ref_atoms = data.get_ref_atoms(
     structures=['bcc*','fcc*','hcp*'],
     quantities=['energy'],
     sort_by='energy')
 ref_data = data.get_ref_data()
\end{verbatim}
}

The user-specified reference data, i.e. structures and properties, are passed to the calculator and optimization kernel, respectively.

\subsection{Initial model: CATParam}
\label{sec:CATParam}

The construction of TB/BOP models in BOPcat relies on iterative optimization steps that require an initial guess that can be provided by the user or generated by the CATParam module. 
In the latter case the functions and parameters are initialized from a recently developed database of TB bond integrals from downfolded DFT eigenspectra~\cite{Jenke2019}. 
In the following snippet, the initial model is read from the filename specified in the input controls.

{\footnotesize
\begin{verbatim}
 from bopcat.catparam import CATParam
 param = CATParam(controls=controls)
 ini_model = param.models[0]
\end{verbatim}
}

The CATParam module also provides meta-operations on sets of models such as identifying the optimum model for a given set of reference data.

\subsection{Calculator: CATCalc}
\label{sec:CATCalc}

After the reference structures are read and the initial model is generated, the CATCalc module sets up the calculator for the computation of the specified properties for the reference structures by the initial and then optimized TB/BOP model. 
To this end, a list of BOPfox-ASE calculators is constructed which can be run in serial or in multiprocessing mode. 
The latter is optimized by a load balancing scheme that is based on the size of the structures. 
Additional properties for the model optimization can easily be included by extending this module accordingly.
In the following snippet, an initial model is provided and an ASE calculator is initialized for each of the reference structures.

{\footnotesize
\begin{verbatim}
 from bopcat.catcalc import CATCalc
 calc = CATCalc(controls=ctr,
                model=ini_model,
                atoms=ref_atoms)
\end{verbatim}
}

Here, the input controls (\verb+ctr+), the initial guess for the TB/BOP model (\verb+ini_model+) and the reference structures (\verb+ref_atoms+) are specified in the code snippets given in Sec.~\ref{sec:CATControls},~\ref{sec:CATParam} and~\ref{sec:CATData}, respectively.

\subsection{Construction and testing kernel: CATKernel}
\label{sec:CATKernel}

The reference data and associated calculators are then used to set up the CATKernel module.
This module provides an interface to the objective function for the construction and testing of TB/BOP models. 
The default objective function for $N_p$ properties of $N_s$ structures is given by
\begin{equation}
\chi^2 = \sum_p \frac{w_p}{N_p} \sum_s \frac{\tilde{w_{ps}}}{N_s^{(p)}} \frac{X_{ps}^\mathrm{model}-X_{ps}^\mathrm{ref}}{\bar{X}^\mathrm{ref}_p} .
\end{equation}
The user can choose other definitions of the objective function that are implemented in BOPcat or provide external implementations.
The weights for the individual structures 
\begin{equation}
 \tilde{w_{ps}} = \frac{1}{\gamma_pN_\mathrm{atoms}^{(i)}}
\end{equation}
are specified by the user or determined by the module. 
The dimensionality factor $\gamma_{p}$ balances the relative weights of energies, forces and stress by taking values of $1,3,6$, respectively. 
The objective function can also be normalized by the variance of the properties as in Ref.~\cite{Krishnapriyan2017}.
For construction purposes, this module provides an interface between objective function and optimizer algorithms that are either readily available in the Python modules Scipy~\cite{Scipy} and NLopt~\cite{Nlopt} or provided by the user as external module. 

In the following example we illustrate the initialization and running of the CATkernel object for the input controls (\verb+ctr+), calculator (\verb+calc+) and reference properties (\verb+ref_data+) defined in Sec.~\ref{sec:CATControls},~\ref{sec:CATCalc} and~\ref{sec:CATParam}.

{\footnotesize
\begin{verbatim}
 from bopcat.catkernel import CATKernel
 kernel = CATKernel(controls=ctr,
                    calc=calc,
                    ref_data)
 kernel.run()                  
\end{verbatim}
}

\subsection{Parallel execution of BOPcat}

The optimization kernel can be executed efficiently in parallel by distributing the required property calculations with Python subprocesses and the message passing interface. 
For this purpose, BOPcat also features a process management module to interact with the queuing system of compute clusters.
This allows high-throughput optimizations of TB/BOP models with, e.g., different initial models or different sets of reference data.
We illustrate this in the following snippet:

{\footnotesize
\begin{verbatim}
 from bopcat.process_management\
 import Process_catkernels
 from bopcat.process_management\
 import Process_catkernel
 from bopcat.process_management import queue
 models = [ini_model.rattle(kernel.variables) 
           for i in range(5)]
 subprocs = []
 for i in range(len(kernels)):
     ckern = kernel.copy()
     ckern.calc.set_model(models[i])
     proc = Process_catkernel(catkernel=ckern,
                              queue=queue)
     subprocs.append(proc)
 proc = Process_catkernels(procs=subprocs)
 proc.run()
\end{verbatim}
}

A list of random models (\verb+models+) is first generated by applying noise to the parameters of the initial model (\verb+ini_model+). 
A copy of the CATKernel object is then generated (\verb+ckern+) and a model from the list is assigned. 
The kernel is then serialized by the \verb+Process_catkernel+ object which contains the details on how to execute the kernel. 
Finally, the individual subprocesses are wrapped in the main object \verb+Process_catkernels+ which submits them to a specified queue (\verb+queue+) of the compute cluster.

\subsection{Graphical user interface}

\begin{figure*}[t]
 \centering
 \includegraphics[width=1.75\columnwidth]{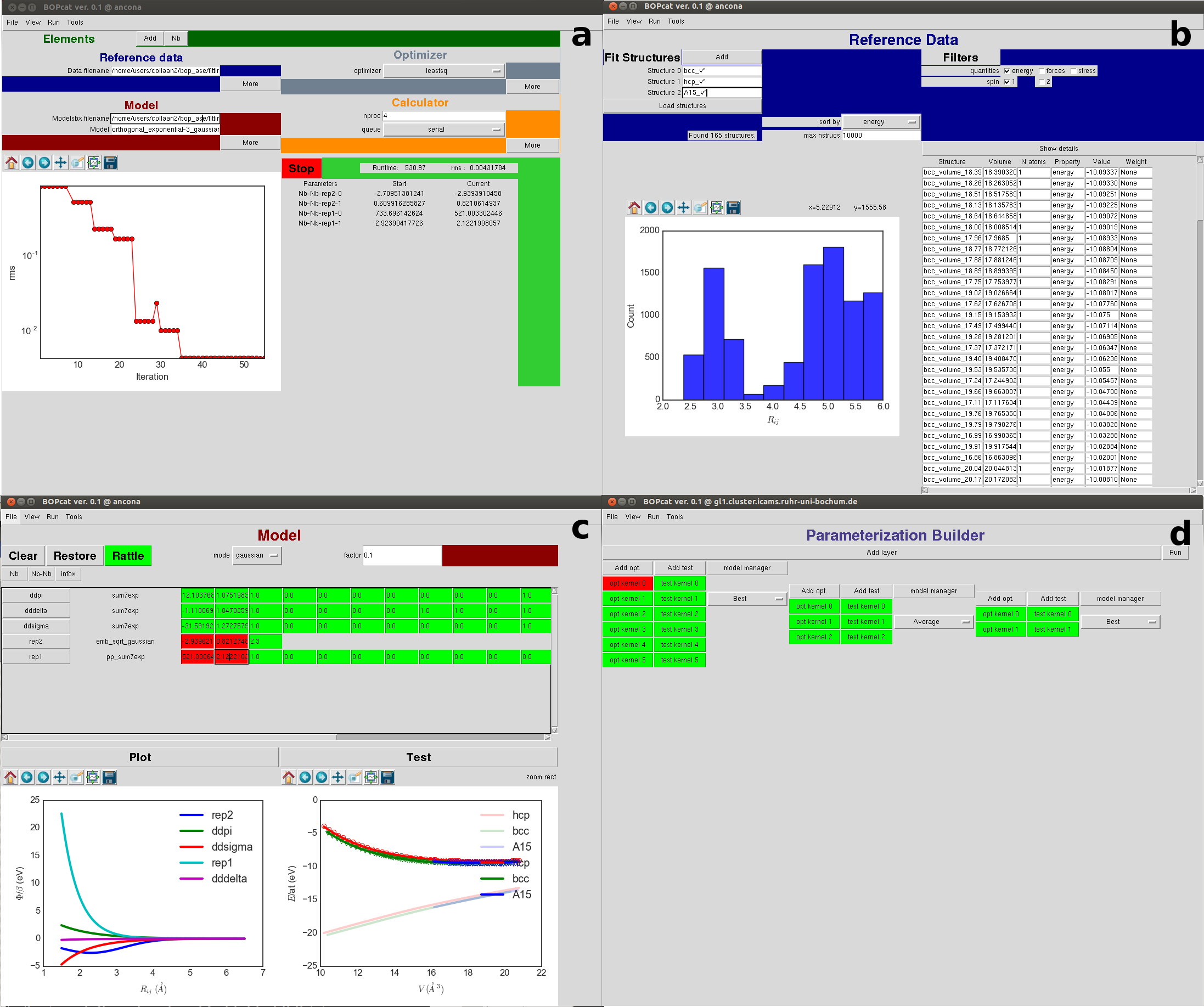}
 \caption{Snapshots of the graphical user interface for (a) constructing a TB/BOP model, (b) generating reference structures,
          (c) setting optimization variables and (d) defining a parameterization protocol.}
 \label{fig:gui}         
\end{figure*}

The setup of a construction or testing process with BOPcat discussed so far is based on writing and editing Python scripts. 
Alternatively, BOPcat can be driven with a graphical user interface (GUI), see snapshots in Fig.~\ref{fig:gui}. 
Basic information of a construction such as the current value of the parameters and objective function can be visualized. 
The reference data can be inspected visually and corresponding attributes such as the weights can be set directly.
For testing purposes, a given TB/BOP model can be evaluated and visualized for a given set of reference data. 
The GUI also comes with a parameterization builder which allows one to design and execute a parameterization protocol. 
In particular, one can add a number of optimization layers that consist of several optimization kernels  with communication of intermediate TB/BOP models between subsequent layers. 
This provides the technical prerequisite for automatizing the construction of TB/BOP models by sophisticated protocols.

\section{Construction and testing of simple Fe BOP}\label{sec:application}

\subsection{Functional form}

As an example application of BOPcat we construct a simple BOP model for Fe including magnetism.
We note that this Fe BOP is constructed to demonstrate BOPcat and that capturing the reference data more precisely would require a more complex functional form.
The atoms are assumed to be charge-neutral so that the binding energy given in Eq.~\ref{eq:U_B_general} reduces to
\begin{equation}
\label{eq:U_B}
 U_\mathrm{binding} = U_\mathrm{bond} + U_\mathrm{rep} + U_\mathrm{X} + U_\mathrm{emb}.
\end{equation}
We choose a $d$-valent model including only two-center contributions, similar to previous TB/BOP models for Fe~\cite{Mrovec2011,Ford2015,Madsen2011}, with an initial number of $d$-electrons of $N_d=6.8$. 
We use a BOP expansion up to the 9-th moment (Eq.~\ref{eq:mu}) as a good compromise between performance and accuracy of the DOS.

The Hamiltonian matrix elements $H_{i\alpha j\beta}$ which enter the calculation of the moments are constructed from Slater-Koster bond integrals $\beta_{ij}^\sigma$ with the angular character  $\sigma=dd\sigma, dd\pi, dd\delta$. 
The initial guess of the bond integral parameters is taken from projections of the DFT eigenspectrum on a TB minimal basis for the Fe-Fe dimer~\cite{Jenke2019}. 
The distance dependence of the bond integrals is represented by a simple exponential function
\begin{equation}
\beta_{ij}^\sigma = a\exp{-bR_{ij}^c} .
\end{equation}
For this simple BOP model, the repulsive energy is given as a simple pairwise contribution.
\begin{equation}
 U_\mathrm{rep} = \frac{1}{2}\sum_{i,i\neq j}a_\mathrm{rep}\exp\left(-b_\mathrm{rep}R_{ij}^{c_\mathrm{rep}}\right).
\end{equation}
The magnetic contribution to the energy $U_X$ is evaluated using
\begin{equation}\label{eq:Stoner}
 U_\mathrm{X} = -\frac{1}{4}\sum_i I m_i^2
\end{equation}
with $m_i=N_i^\uparrow-N_i^\downarrow$ the magnetic moment of atom $i$ and $I$ the Stoner exchange integral that is initially set to 0.80~eV similar to Ref.~\cite{Mrovec2011}.
In extension to Eq.~\ref{eq:U_B_general} we use an additional empirical embedding contribution in Eq.~\ref{eq:U_B}
\begin{equation}
 U_\mathrm{emb} = -\sum_i\sqrt{\sum_{j,j\neq i} a_\mathrm{emb}\exp\left(-b_\mathrm{emb}(R_{ij}-c_\mathrm{emb})^2\right)}
\end{equation}
which may be understood as providing contributions due to the missing $s$ electrons and non-linear exchange correlation.

The bond integrals, repulsive and embedding functions are multiplied with a cut-off function
\begin{equation}
 f_\mathrm{cut} = \frac{1}{2}\left[ \cos\left(\pi \frac{R_{ij}-(r_\mathrm{cut}-d_\mathrm{cut})}{d_\mathrm{cut}}\right) + 1 \right]
\end{equation}
in the range of $[r_\mathrm{cut} - d_\mathrm{cut}$, $r_\mathrm{cut}]$ in order to restrict the range of the interatomic interaction.
For the Fe BOP constructed here we used $r_\mathrm{cut}=3.8~\AA$, $d_\mathrm{cut}=0.5~\AA$ for the bond integrals and $r_\mathrm{cut}=6.0~\AA$, $d_\mathrm{cut}=1.0~\AA$ for the repulsive and embedding energy terms.

\subsection{Reference data}

The reference data for constructing and testing the Fe BOP is obtained by DFT calculations using the \texttt{VASP} software~\cite{Kresse-96-1,Kresse-96-2} with a high-throughput environment~\cite{Hammerschmidt-13}.
We used the PBE exchange-correlation functional~\cite{Perdew-96} and PAW pseudopotentials~\cite{Bloechl-94} with $p$ semicore states. 
A planewave cut-off energy of 450~eV and Monkhorst-Pack $k$-point meshes~\cite{Monkhorst-76} with a density of 6~/\AA$^{-1}$ were sufficient to converge the total energies to 1~meV/atom. 
The reference data for constructing the Fe BOP comprised the energy-volume curves of the bulk structures bcc, fcc, hcp, and the topologically close-packed (TCP) phases A15, $\sigma$, $\chi$, $\mu$, C14, C15 and C36 that are relevant for Fe compounds~\cite{Ladines2015}. 

For testing the Fe BOP, several additional properties were determined that are related to the ferromagnetic bcc groundstate structure. 
In particular, the elastic constants at the equilibrium volume were determined by fitting the energies as function of the relevant strains~\cite{Golesorkhtabar2013}. 
The formation energies of point defects were calculated with the supercell approach with fixed volume corresponding to the bulk equilibrium volume. 
A $6\times 6\times 6$ supercell was found to be sufficient to converge the formation energies to an uncertainty of 0.1~eV.
The energy barriers for vacancy migration were calculated with the nudged elastic band method.
The phonon spectra are computed with the Phonopy software~\cite{Togo2015}. 
The transformation paths from bcc to other crystal structures were determined according to Ref.~\cite{Paidar1999}. 

\subsection{Construction}

The parameters of the Fe BOP with the functions and reference data given above were optimized with the following BOPcat protocol: 
\begin{enumerate}
 \item The magnitudes of the repulsive and embedding terms $a_\mathrm{rep}$ and $a_\mathrm{emb}$ are adjusted to reproduce the energies of hydrostatically deformed bcc, fcc and hcp with fixed values for the exponents taken from Ref.~\cite{Madsen2011}.
 \item The prefactors of the bond integrals are adjusted including the energies of the TCP phases.
 \item The exponents of the repulsive functions are optimized including the randomly deformed structures in the reference set.
 \item The exponents of the bond integrals are optimized by increasing the size of the reference data. 
 \item The other parameters are optimized further while increasing the number of reference structures.
\end{enumerate}

\begin{figure}[htb]
 \centering
 \includegraphics[width=0.99\columnwidth]{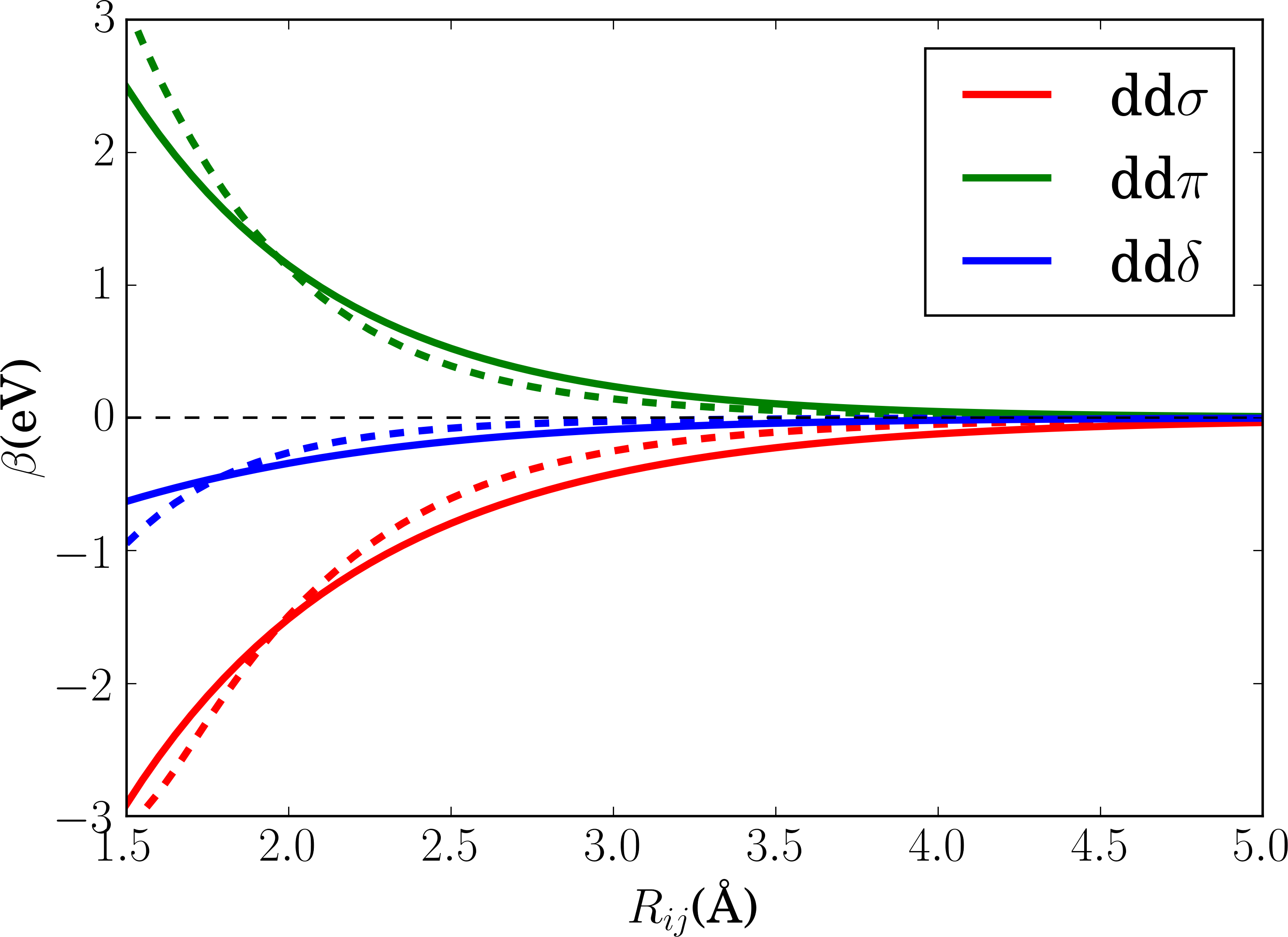}
 \caption{Distance dependence of the $d$ bond integrals as obtained by downfolding from DFT eigenspectra for the Fe-Fe dimer (dashed)~\cite{Jenke2019} and after optimization to Fe bulk structures (solid).}
 \label{fig:bondint_fe}         
\end{figure}

\begin{figure}[htb]
 \centering
 \includegraphics[width=0.99\columnwidth]{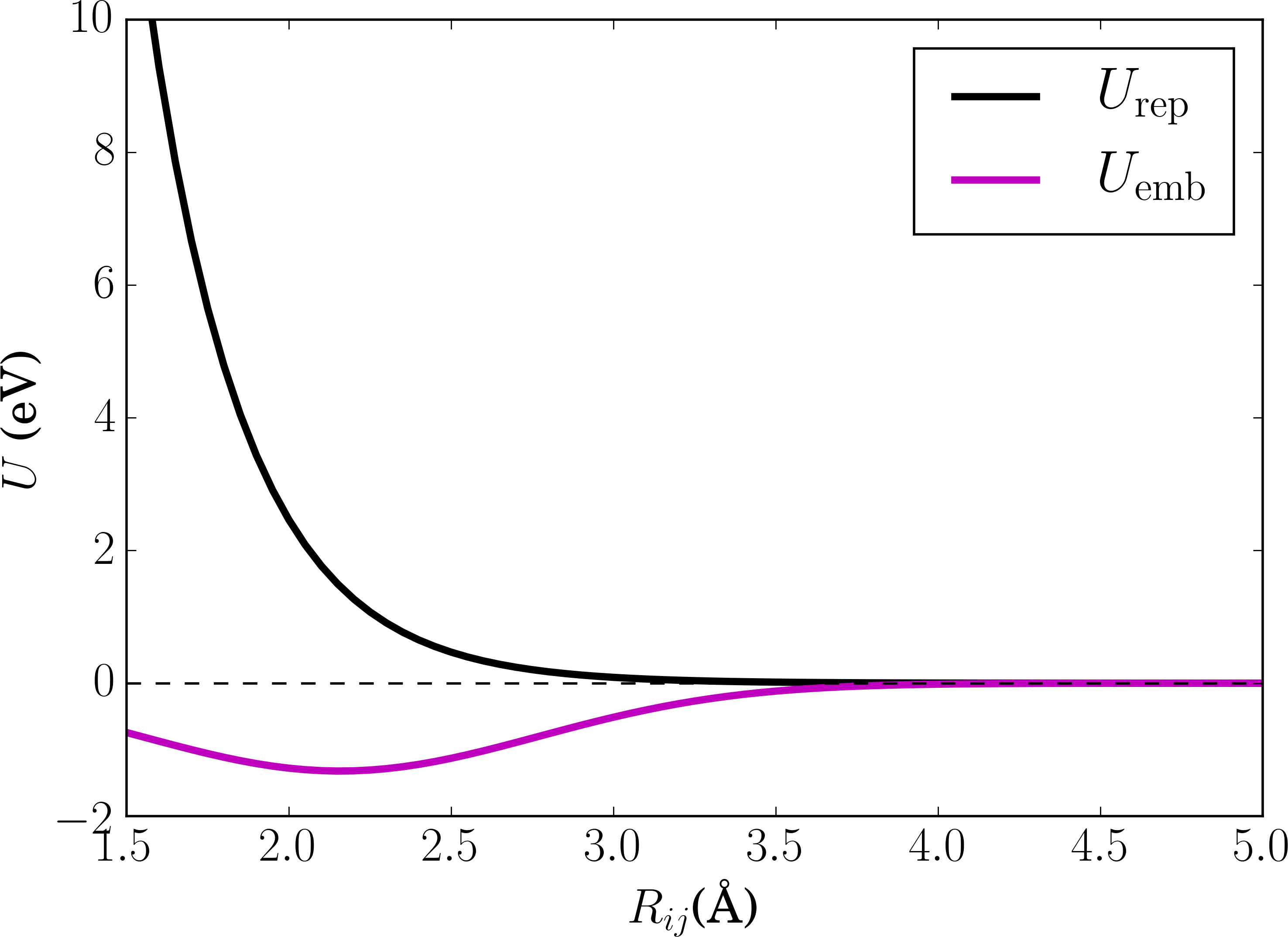}
 \caption{Distance dependence of the pairwise repulsive potential $U_{\mathrm {rep}}$ (black) and the embedding term $U_{\mathrm {emb}}$ (violet) after optimization to Fe bulk structures.}
 \label{fig:U_rep}         
\end{figure}

The resulting parameters of the model are compiled in Tab.~\ref{tab:parameters}.
\begin{table}[t]
 \centering
 \caption{Parameters of the BOP model for Fe. The unit of $b_\mathrm{emb}$ is $\AA^{-2}$ and $c_\mathrm{emb}$ is $\AA$.}
 \scalebox{1.00}{
 \begin{tabular}{l c c c}
 \hline
 \hline
                          & $a$ (eV)     & $b$ ($\AA^{-1}$)    & c       \\
 \hline  
 $dd\sigma$               & -24.9657     & 1.4762              & 0.9253  \\
 $dd\pi$                  &  21.7965     & 1.4101              & 1.0621  \\
 $dd\delta$               &  -2.3536     & 0.7706              & 1.3217  \\
 \hline
 $U_\mathrm{rep}$         & 1797.4946    & 3.2809              & 1.0067  \\
 \hline
 $U_\mathrm{emb}$         & -1.3225      & 1.3374              & 2.1572  \\
 \hline
 $N_d$                    &  6.8876      &                     &         \\
 $I (eV)$                 &  0.9994      &                     &         \\
 \hline
 \hline
 \end{tabular}}
 \label{tab:parameters}
\end{table}
The initial and optimized bond integrals, and the empirical potentials are plotted in Fig.~\ref{fig:bondint_fe}. 
There is no substantial change in the bond integrals except that the bond integrals become longer-ranged after optimization. 
At shorter distance the bond integrals become weaker which can be rationalized by the screening influence of the neighboring atoms in the bulk structure. 
The effective number of $d$ electrons and the Stoner integral increased during optimization.
The optimized pairwise repulsive term $U_{\mathrm{rep}}$ shown in Fig.~\ref{fig:U_rep} is repulsive for all distances. 
The embedding term $U_{\mathrm{emb}}$, in contrast, is negative for all distances.

\subsection{Testing}

\subsubsection{Properties related to bcc-Fe}

An important basic test of the Fe BOP is the performance for properties that are closely related to the ferromagnetic bcc groundstate structure at 0~K. 
In Tab.~\ref{tab:bulk_properties} we compare elastic constants as well as point defect and surface properties as predicted by the Fe BOP to available DFT and experimental values.
The overall good agreement demonstrates the robust transferability of the Fe BOP to properties not included in the training set. 
\begin{table}[htb]
 \centering
 \caption{Bulk properties of ferromagnetic bcc-Fe. The experimental values for the elastic constants, vacancy formation energy and vacancy migration energy are taken from Refs.~\cite{Rayne1961}, ~\cite{DeSchepper1983} and ~\cite{Vehanen1982}, respectively. The DFT values for the vacancy migration energy and surface energies are taken from Refs. ~\cite{Domain2001} and ~\cite{Blonski2007}, respectively.}
 \scalebox{1.00}{
 \begin{tabular}{l c c c}
 \hline
 \hline
                            & BOP     & DFT    & experiment     \\
 \hline  
 V/atom (\AA$^3$)           & 11.48   & 11.46  & 11.70          \\
 $B$ (GPa)                  & 171     & 176    & 168            \\
 $C_{11}$ (GPa)             & 265     & 257    & 243            \\
 $C_{12}$ (GPa)             & 125     & 154    & 138            \\
 $C_{44}$ (GPa)             &  87     & 85     & 122            \\
 $E_f^\mathrm{vac}$ (eV)    & 2.03    & 2.20   & 2.00           \\
 $E_\mathrm{mig}^\mathrm{vac}$ (eV)    & 1.33    & 0.65   & 0.55    \\
 $E_f^\mathrm{100}$ (eV)    & 3.65    & 4.64   &                \\
 $E_f^\mathrm{110}$ (eV)    & 3.13    & 3.64   &                \\
 $E_f^\mathrm{111}$ (eV)    & 3.59    & 4.34   &                \\ 
 $\gamma_{(100)}$ (J/m$^2$) & 1.44    & 2.47 &                    \\
 $\gamma_{(110)}$ (J/m$^2$) & 1.27    & 2.37 &                    \\
 $\gamma_{(111)}$ (J/m$^2$) & 2.04    & 2.58 &                    \\
 $\gamma_{(211)}$ (J/m$^2$) & 1.50    & 2.50 &                    \\
 \hline
 \hline
 \end{tabular}}
 \label{tab:bulk_properties}
\end{table}

The predicted elastic constants $C_{11}$, $C_{12}$ and $C_{44}$ are in good agreement with DFT although the specific deformed structures were not included in the reference set for constructing the BOP. 
The prediction for the vacancy formation energy $E_f^\mathrm{vac}$ is also in good agreement with DFT while the vacancy migration energy $E_\mathrm{mig}^\mathrm{vac}$ is overestimated as in previous TB/BOP models for Fe~\cite{Madsen2011, Mrovec2011}.
The formation energies of the vacancy and self-interstitial atoms calculated by the Fe BOP exhibit the correct energetic ordering although they were not included in the parameterization. 
The absolute values are slightly underestimated as compared to DFT, similar to previous models~\cite{Madsen2011, Mrovec2011}.
The relative stability of the low-index surfaces are also satisfactorily reproduced by the present Fe BOP. However, similar to the previous BOP model of Mrovec\cite{Mrovec2011}, the energy difference of the other surfaces relative to $(110)$ is overestimated.
The deviations for point defects and surfaces are attributed to the missing $s$ electrons in the model and the lack of screening in the orthogonal TB model.

\begin{figure}[htb!]
 \centering
 \includegraphics[width=0.99\columnwidth]{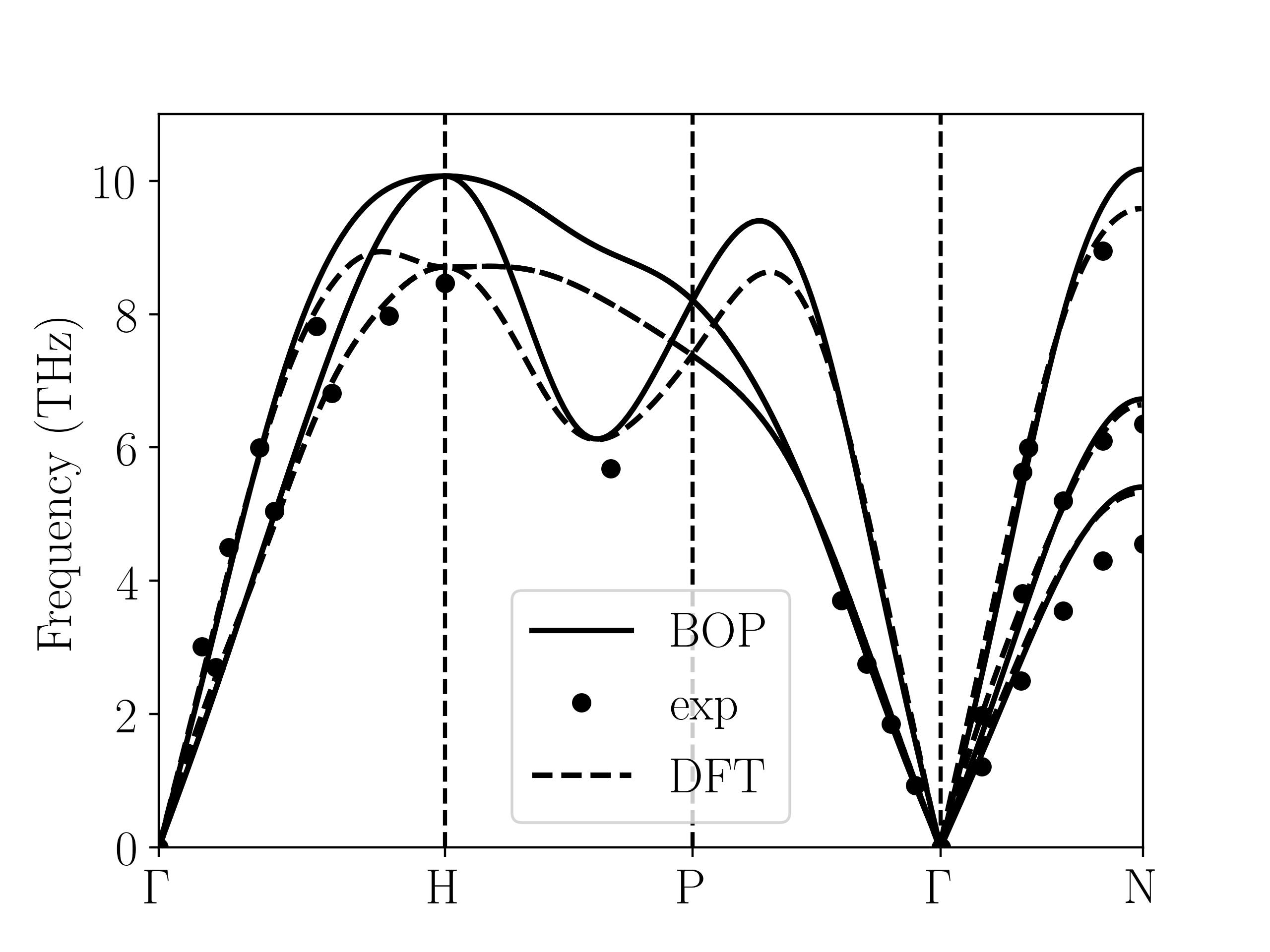}
 \caption{Phonon bandstructure of ferromagnetic bcc-Fe. The solid (dashed) lines are the BOP (DFT) results
          while the symbols correspond to experimental data taken from Ref~\cite{Klotz2000}.}
 \label{fig:phonon_fe}         
\end{figure}

As a test towards finite-temperature applications we determine the phonon bandstructure for ferromagnetic bcc Fe shown in Fig.~\ref{fig:phonon_fe}.
The prediction of the Fe BOP is in good overall agreement with DFT and experimental results. 
The close match near the $\Gamma$ point is expected from the good agreement of the elastic constants. 
The experimental data were obtained at 300~K which can explain the deviation of the experimental from the
DFT/BOP data for the $T2$ branch along $T-N$.  

\begin{figure}[h]
 \centering
 \includegraphics[width=0.99\columnwidth]{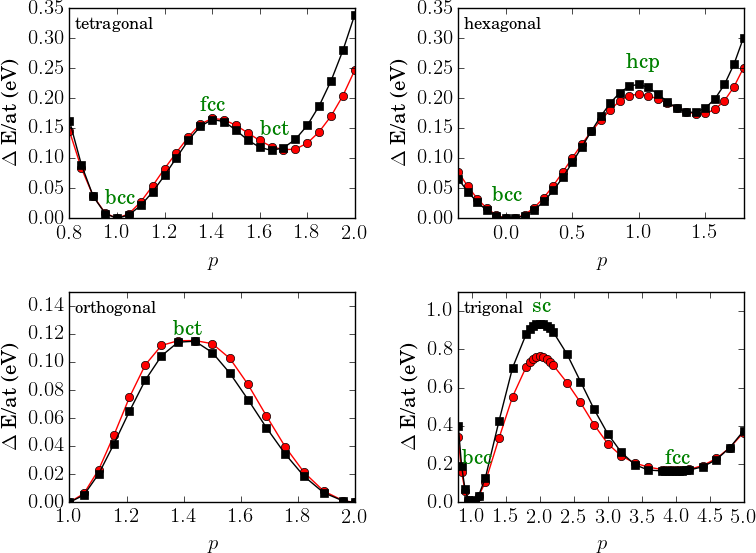}
 \caption{Energy profile along the transformation path connecting the ferromagnetic bcc Fe with other high-symmetry structures as obtained by the Fe BOP (red) and DFT (black).}
 \label{fig:trans_fe}         
\end{figure}

The transferability of the model in the case of large deformations of the bcc groundstate structure is tested using deformation paths that connect the high symmetry structures bcc, fcc, hcp, sc and bct.
The energy profile versus the deformation parameter for the tetragonal, hexagonal, orthogonal and trigonal deformation paths is shown in Fig.~\ref{fig:trans_fe}. 
The energy profiles for all transformation paths are predicted very well by the Fe BOP model. 
The energies at the high symmetry points are in good agreement with DFT except for the sc structure which is slightly underestimated by BOP. 

\subsubsection{Transferability to other crystal structures}

\begin{figure}[htb!]
 \centering
 \includegraphics[width=0.99\columnwidth]{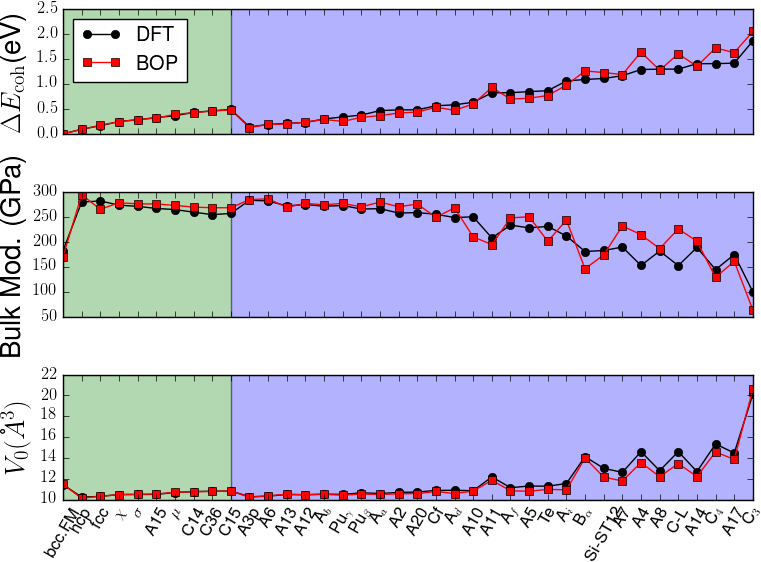}
 \caption{Relative binding energy, bulk modulus and equilibrium volume of bulk structures including TCP phases. The difference in background shading indicates structures that belong to the reference set used for construction (green) and for testing (blue).}
 \label{fig:eos_fe}         
\end{figure}

In order to verify the transferability of the Fe BOP model to other crystal structures, we determined the equilibrium binding energy, volume and bulk modulus for a broader set of crystal structures as shown in Fig.~\ref{fig:eos_fe}. 
The properties of the bcc, fcc, hcp, A15, $\sigma$, $\chi$, $\mu$, C14, C15 and C36 structures that were included in the parameterization are reproduced with excellent agreement.
This shows that the chosen functional form of the Fe BOP is sufficiently flexible to adapt to this set of reference data.
The Fe BOP model also shows robust predictions of equilibrium binding energy, volume and bulk modulus across the set of tested structures.
The open crystal structures (e.g. A4) show comparably larger errors which can be expected as the reference set used in the parameterization covers mostly close-packed local atomic environments~\cite{Jenke2018} while local atomic environments of open structures were not part of the training set.

\section{Conclusions}

We present the BOPcat software for the construction and testing of TB/BOP models as implemented in the BOPfox code. 
TB/BOP models are parameterized to reproduce reference data from DFT calculations including energies, forces, and stresses as well as derived properties like defect formation energies.
The modular framework of BOPcat allows one to implement complex parameterization protocols by flexible python scripts. 
BOPcat provides a graphical user interface and a highly-parallelized optimization kernel for fast and efficient handling of large data sets. 

We illustrate the key features of the BOPcat software by constructing and testing a simple $d$-valent BOP model for Fe including magnetism. 
The resulting BOP predicts a variety of properties of the groundstate structure with good accuracy and shows good quantitative transferability to other crystal structures.

\section*{Acknowledgements}

The authors are grateful to Aparna Subramanyam, Malte Schr{\"o}der, Alberto Ferrari, Jan Jenke, Yury Lysogorskiy, Miroslav Cak, and Matous Mrovec for discussions and feedback using BOPcat.
We acknowledge financial support by the German Research Foundation (DFG) through research grant HA 6047/4-1 and project C1 of the collaborative research centre SFB/TR 103.
Part of the work was carried out in the framework of the International Max-Planck Research School SurMat.

\appendix
\section*{Appendix}
We provide several examples to illustrate the script required
to direct an optimization procedure. The examples show only the 
primary capabilities of BOPcat. For a complete list of possible
functionalities of the software, the user is referred to the 
user manual.

\section{Constructing Fe-Nb EAM}
Aside from parameterizing TB models and BOPs, the software is also
capable of constructing empirical potentials. We illustrate
this in the following script which can be used to parameterize
an EAM for the Fe-Nb binary. The model consists of pair potentials
described by polynomials, and square root embedding functions. 
The optimization proceeds in four steps, each parameterizing
different interaction pairs (1) Fe-Fe, (2) Nb-Nb, (3) Fe-Nb and (4) all.
In the first three steps, the reference data consists only of energies,
while the last step include structures with energy, forces and stress. 

{\scriptsize
\begin{verbatim}
from bopcat.catcontrols import CATControls
from bopcat.catdata import CATData
from bopcat.catparam import CATParam
from bopcat.catcalc import CATCalc
from bopcat.catkernel import CATKernel
import bopcat.functions as Funcs

#################### Input controls ####################
cat_controls = CATControls()
## required, list of elements
cat_controls.set(elements = ['Fe','Nb'])
# Model
## cut_offs
cat_controls.set(model_cutoff =\
    {'r2cut': {'Fe-Fe':6.0, 'Nb-Nb':6.0, 'Fe-Nb':6.0}
    ,'d2cut': {'Fe-Fe':0.5, 'Nb-Nb':0.5, 'Fe-Nb':0.5}})
## functional forms    
cat_controls.set(model_functions =\
    {'rep1':{'Fe-Fe':Funcs.polynom4()
            ,'Nb-Nb':Funcs.polynom4()
            ,'Fe-Nb':Funcs.polynom4()}
    ,'rep2':{'Fe-Fe':Funcs.sqrt_polynom3
            ,'Nb-Nb':Funcs.sqrt_polynom3
            ,'Fe-Nb':Funcs.sqrt_polynom3}})        
# Calculator
## dictionary of keyword-values used by calculator 
cat_controls.set(calculator_settings =\
    {'version':'eam'})
# Reference Data
## database filters
cat_controls.set(data_parameters =\
    {'xc_functional':'PBE','encut':450})
## filename
cat_controls.set(data_filename = 'FeNb.fit')
# optimizer
cat_controls.set(opt_optimizer_options =\
    {'max_nfev':100,'diff_step':0.01}) 
cat_controls.set(opt_optimizer = 'least_squares')
# others
cat_controls.set(verbose = 2)

#################### Initialization ####################
# generate reference data
cat_data = CATData(controls=cat_controls)
# generate initial model
cat_param = CATParam(controls=cat_controls)
# set up calculator
cat_calc =\
    CATCalc(controls=cat_controls
           ,model=cat_param.models[-1])

################### Optimization Fe ####################
# optimize Fe
## fetch reference structures and properties
## structures can be identified by structure name or 
## by 'stoich/sgrp/sys-type/calc-type/calc-order'
ref_atoms =\
    cat_data.get_ref_atoms(structures=['Fe*/*/0/*/*']
                          ,quantities=['energy'])
ref_data = cat_data.get_ref_data()
cat_calc.set_atoms(ref_atoms)
## optimization variables
var = [{"bond":["Fe","Fe"],'rep1':[True,True,True,True]
       ,'rep2':[True,True,True]}]
## set up optimization kernel 
optfunc = CATKernel(calc=cat_calc,ref_data=ref_data
                   ,variables=var,log='log.cat'
                   ,controls=cat_controls)
# run optimization
optfunc.optimize()
# update models
new_model = optfunc.optimized_model
cat_param.models.append(new_model)

################### Optimization Nb ####################
## start from previous model
cat_calc.set_model(new_model)
ref_atoms =\
    cat_data.get_ref_atoms(structures=['Nb*/*/0/*/*']
                          ,quantities=['energy'])
ref_data = cat_data.get_ref_data()
cat_calc.set_atoms(ref_atoms)
var = [{"bond":["Nb","Nb"],'rep1':[True,True,True,True]
       ,'rep2':[True,True,True]}]
optfunc = CATKernel(calc=cat_calc,ref_data=ref_data
                   ,variables=var,log='log.cat'
                   ,controls=cat_controls)
optfunc.optimize()
new_model = optfunc.optimized_model
cat_param.models.append(new_model)

################### Optimization Fe-Nb ###################
## start from previous model
cat_calc.set_model(new_model)
ref_atoms =\
    cat_data.get_ref_atoms(structures=['Fe*Nb*/*/0/*/*']
                          ,quantities=['energy'])
ref_data = cat_data.get_ref_data()
cat_calc.set_atoms(ref_atoms)
var = [{"bond":["Fe","Nb"],'rep1':[True,True,True,True]
       ,'rep2':[True,True,True]}]
optfunc = CATKernel(calc=cat_calc,ref_data=ref_data
                   ,variables=var,log='log.cat'
                   ,controls=cat_controls)
optfunc.optimize()
new_model = optfunc.optimized_model
cat_param.models.append(new_model)

#################### Optimization all ####################
## start from previous model
cat_calc.set_model(new_model)
ref_atoms =\
    cat_data.get_ref_atoms(structures=['Fe*Nb*/*/0/*/*']
                          ,quantities=['energy','forces'
                                      ,'stress'])
ref_data = cat_data.get_ref_data()
cat_calc.set_atoms(ref_atoms)
var = [{"bond":["Fe","Fe"],'rep1':[True,True,True,True]
       ,'rep2':[True,True,True]}
      ,{"bond":["Nb","Nb"],'rep1':[True,True,True,True]
       ,'rep2':[True,True,True]}
      ,{"bond":["Fe","Nb"],'rep1':[True,True,True,True]
       ,'rep2':[True,True,True]}]
optfunc = CATKernel(calc=cat_calc,ref_data=ref_data
                   ,variables=var,log='log.cat'
                   ,controls=cat_controls)
optfunc.optimize()
new_model = optfunc.optimized_model
cat_param.models.append(new_model)
# write out new model
new_model.write(filename='models_optimized.bx')
\end{verbatim}
}

\section{Construction of BOP from raw TB bond integrals}\label{sec:app2}
A BOP or TB model can also be constructed directly from the numerical data 
for the bond integrals using BOPcat. One simply needs to supply the 
function to fit the data. In the following example, a BOP
is constructed for Fe. The electronic 
part of the potential is parameterized with respect to 
the DFT eigenvalues while the other terms are optimized
by fitting to the energies of the bcc structures.

{\scriptsize
\begin{verbatim}
from bopcat.catcontrols import CATControls
from bopcat.catdata import CATData
from bopcat.catparam import CATParam
from bopcat.catcalc import CATCalc
from bopcat.catkernel import CATKernel
import bopcat.functions as Funcs
from bopcat.eigs import get_relevant_bands

#################### Input controls ####################
cat_controls = CATControls()
## required, list of elements
cat_controls.set(elements = ['Fe'])
# Model
## cut_offs
cat_controls.set(model_cutoff =\
{'rcut' : {'Fe-Fe':4.0},'r2cut': {'Fe-Fe':6.0}
,'dcut' : {'Fe-Fe':0.5},'d2cut': {'Fe-Fe':0.5}})
## TB basis
cat_controls.set(model_orthogonal = True)
cat_controls.set(model_valences={'Fe':'d'})
cat_controls.set(model_betabasis='tz0')
cat_controls.set(model_betatype='loewdin')
## filenames to bond integrals and onsites
cat_controls.set(model_pathtobetas='../betas')
cat_controls.set(model_pathtoonsites='../onsites')
## functional forms
cat_controls.set(model_functions =\
    {'ddsigma':{'Fe-Fe':Funcs.exponential()}
    ,'ddpi':{'Fe-Fe':Funcs.exponential()}
    ,'dddelta':{'Fe-Fe':Funcs.exponential()}
    ,'rep1':{'Fe-Fe':Funcs.exponential()}
    ,'rep2':{'Fe-Fe':Funcs.sqrt_gaussian()}})
# Calculator
## dictionary of keyword-values used by calculator 
cat_controls.set(calculator_settings =\
    {'bandwidth':'findeminemax'
    ,'terminator':'constantabn','bopkernel':'jackson'
    ,'nexpmoments':100,'moments':9,'version':'bop'})

# Reference Data
## database filters
cat_controls.set(data_parameters =\
    {'xc_functional':'PBE','encut':450})
## filename
cat_controls.set(data_filename = 'Fe.fit')
# optimizer
cat_controls.set(opt_optimizer_options =\
    {'max_nfev':100,'diff_step':0.01}) 
cat_controls.set(opt_optimizer = 'least_squares')
# others
cat_controls.set(verbose = 2)

#################### Initialization ####################
# generate reference data
cat_data = CATData(controls=cat_controls)
# generate initial model
cat_param = CATParam(controls=cat_controls)
# set up calculator
cat_calc =\
    CATCalc(controls=cat_controls
           ,model=cat_param.models[-1])

################## Optimization elec ###################
ref_atoms =\
    cat_data.get_ref_atoms(structures=['Fe/229/0/0/*']
                          ,quantities=['eigenvalues'])
ref_data = cat_data.get_ref_data()
orb_char = cat_data.orb_char
cat_calc.set_atoms(ref_atoms)
# select only relevant bands in dft eigs
ref_data, orb_char =\
    get_relevant_bands(cat_calc,ref_data,orb_char)
# optimize bond integrals
var  = [{"bond":["Fe","Fe"],"ddsigma":[True,True]
        ,"ddpi":[True,True], "dddelta":[True,True]
        ,'atom':'Fe','onsitelevels':[True]}]
# set up optimization 
optfunc = CATKernel(calc=cat_calc,ref_data=ref_data
                   ,variables=var,log='log.cat'
                   ,controls=cat_controls
                   ,weights=orb_char)
# run optimization
optfunc.optimize()
# update models
new_model = optfunc.optimized_model
cat_param.models.append(new_model)

################## Optimization rep ####################
ref_atoms =\
    cat_data.get_ref_atoms(structures=['Fe/229/0/1/*']
                          ,quantities=['energy'
                                      ,'vacancy_energy']
                                      )
ref_data = cat_data.get_ref_data()
cat_calc.set_atoms(ref_atoms)
# start from previous model
cat_calc.set_model(new_model)
# optimization variables
var = [{"bond":["Fe","Fe"],'rep1':[True,True]
       ,'rep2':[True,True]}]
# set up optimization 
optfunc = CATKernel(calc=cat_calc,ref_data=ref_data
                   ,variables=var,log='log.cat'
                   ,controls=cat_controls)
# run optimization
optfunc.optimize()
# update models
new_model = optfunc.get_optimized_model()
cat_param.models.append(new_model)
# write out new model
new_model.write(filename='models_optimized.bx')
\end{verbatim}
}

\section{Multiprocessing and usage on clusters}
In order to speed up the optimization process, BOPcat can also
be executed on clusters. The following script illustrates 
this by optimizing a BOP for Re with respect to different sets of 
reference data. A kernel for each case is generated and serialized
by the Process\_catkernel object. These subprocesses are then wraped in a
main process using the Process\_catkernels object which facilitates
the submission to the queing system. A unique queuing system can be easily set up by extending
the \verb+queue+ object.  

{\scriptsize
\begin{verbatim}
from bopcat.catcontrols import CATControls
from bopcat.catdata import CATData
from bopcat.catparam import CATParam
from bopcat.catcalc import CATCalc
from bopcat.catkernel import CATKernel
from process_management import Process_catkernel,vulcan
from process_management import Process_catkernels

#################### Input controls ####################
cat_controls.set(elements = ['Re'])
# model name
cat_controls.set(model =\
    "Cak-2014")
cat_controls.set(model_pathtomodels = 'models.bx')
cat_controls.set(calculator_settings = {'scfsteps':100})
cat_controls.set(data_parameters =\
    {'xc_functional':'PBE','encut':450, 'spin':1})
# reference data
cat_controls.set(data_filename = 'Re.fit')
# free atom energies 
cat_controls.set(data_free_atom_energies = {'Re':-0.5})
# optimizer
cat_controls.set(opt_optimizer_options = {'maxiter':100})
cat_controls.set(opt_optimizer = 'nelder-mead')
# others
cat_controls.set(calculator_nproc = 16)
cat_controls.set(calculator_parallel = 'multiprocessing')
cat_controls.set(verbose = 2)

#################### Initialization ####################
# generate reference data
cat_data = CATData(controls=cat_controls)
# generate initial model
cat_param = CATParam(controls=cat_controls)
# set up calculator
cat_calc =\
    CATCalc(controls=cat_controls
           ,model=cat_param.models[-1])
           
####################### Kernels #########################
var = [{"bond":["Re","Re"],'rep1':[True,False,False,True
      ,False,False,False,False],'rep2':[True,True,True
      ,True,True,False,False,False,False]},{"atom":"Re"
      ,'valenceelectrons':[True]}]
# use different reference structures for each kernel
strucs_list = [['bcc*','hcp*'],['fcc*','hcp*']
              ,['A15*','hcp*'],['C14*','hcp*']]
kernels = []
for i in range(len(strucs_list)):
    ref_atoms =\
    cat_data.get_ref_atoms(structures=strucs_list[i]
                          ,quantities=['energy'])
    ref_data = cat_data.get_ref_data()
    cat_calc.set_atoms(ref_atoms)
    kern = CATKernel(calc=cat_calc,ref_data=ref_data
                   ,variables=var,log='log_\%d.cat'\%i
                   ,controls=cat_controls)
    kernels.append(kern)

####################### Queues ##########################    
queue = vulcan(cores=16,qname='parallel16.q'
              ,pe='mpi16',runtime=60*60*24*14)
subprocs = []
for kern in kernels:
    subprocs.append(Process_catkernel(catkernel=kern
    ,queue=queue,directives=['source ~/.bashrc']))

proc = Process_catkernels(procs=subprocs)  
proc.run()
res = proc.get_results(wait=True)
for i in range(len(proc._procs)):
    kern = proc._procs[i].get_kernel()._catkernel
    opt_model - kern.get_optimized_model()
    cat_param.models.append(opt_model)
    
ave_model = cat_param.average_modelsbx()
ave_model.write(filename='models_optimized.bx')
proc.clean()    
\end{verbatim}
}

\bibliographystyle{unsrt}
\bibliography{bopcat}

\begin{thebibliography}{10}

\bibitem{Drautz2006}
R.~Drautz and D.~G. Pettifor.
\newblock Valence-dependent analytic bond-order potential for transition
  metals.
\newblock {\em Phys Rev B}, 74:174117, 2006.

\bibitem{Drautz2011}
R.~Drautz and D.~G. Pettifor.
\newblock Valence-dependent analytic bond-order potential for magnetic
  transition metals.
\newblock {\em Phys Rev B}, 84:214114, 2011.

\bibitem{Sutton-88}
A.~P. Sutton, M.~W. Finnis, D.~G. Pettifor, and Y.~Ohta.
\newblock The tight-binding bond model.
\newblock {\em J Phys C Solid State}, 21:35, 1988.

\bibitem{Hammerschmidt-09-IJMR}
T.~Hammerschmidt, R.~Drautz, and D.~G. Pettifor.
\newblock Atomistic modelling of materials with bond-order potentials.
\newblock {\em Int J Mater Res}, 100:1479, 2009.

\bibitem{Drautz2015}
R.~Drautz, T.~Hammerschmidt, M.~{\v{C}}{\'{a}}k, and D.~G. Pettifor.
\newblock Bond-order potentials: derivation and parameterization for refractory
  elements.
\newblock {\em Model Simul Mater Sc}, 23(7):074004, 2015.

\bibitem{Mrovec2004}
M.~Mrovec, D.~Nguyen-Manh, D.~G. Pettifor, and V.~Vitek.
\newblock Bond-order potential for molybdenum: Application to dislocation
  behavior.
\newblock {\em Phys Rev B}, 69:094115, 2004.

\bibitem{Mrovec2007}
M.~Mrovec, R.~Gr\"oger, A.~G. Bailey, D.~Nguyen-Manh, C.~Els\"asser, and
  V.~Vitek.
\newblock Bond-order potential for simulations of extended defects in tungsten.
\newblock {\em Phys Rev B}, 75:104119, 2007.

\bibitem{Mrovec2011}
M.~Mrovec, D.~Nguyen-Manh, C.~Els\"asser, and P.~Gumbsch.
\newblock Magnetic bond-order potential for iron.
\newblock {\em Phys Rev Lett}, 106:246402, 2011.

\bibitem{Seiser-11-2}
B.~Seiser, T.~Hammerschmidt, A.~N. Kolmogorov, R.~Drautz, and D.~G. Pettifor.
\newblock Theory of structural trends within 4d and 5d transition metals
  topologically close-packed phases.
\newblock {\em Phys Rev B}, 83:224116, 2011.

\bibitem{Cak2014}
M.~{\v{C}}{\'{a}}k, T.~Hammerschmidt, J.~Rogal, V.~Vitek, and R.~Drautz.
\newblock Analytic bond-order potentials for the bcc refractory metals {Nb},
  {Ta}, {Mo} and {W}.
\newblock {\em J Phys Condens Mat}, 26(19):195501, 2014.

\bibitem{Ford-14}
M.E. Ford, R.~Drautz, T.~Hammerschmidt, and D.~G. Pettifor.
\newblock Convergence of an analytic bond-order potential for collinear
  magnetism in {Fe}.
\newblock {\em Model Simul Mater Sc}, 22:034005, 2014.

\bibitem{Ford2015}
M.~E. Ford, D.~G. Pettifor, and R.~Drautz.
\newblock Non-collinear magnetism with analytic bond-order potentials.
\newblock {\em J Phys Condens Mat}, 27(8):086002, 2015.

\bibitem{Wang-19}
N.~Wang, T.~Hammerschmidt, J.~Rogal, and R.~Drautz.
\newblock Accelerating spin-space sampling by auxiliary spin dynamics and
  temperature-dependent spin-cluster expansion.
\newblock {\em Phys Rev B}, 99:094402, 2019.

\bibitem{Brommer2007}
P.~Brommer and F.~Gähler.
\newblock Potfit: effective potentials from ab initio data.
\newblock {\em Model Simul Mater Sc}, 15(3):295, 2007.

\bibitem{Duff2015}
A.~I. Duff, M.W. Finnis, P.~Maugis, B.~J. Thijsse, and M.~H.F. Sluiter.
\newblock {MEAMfit}: A reference-free modified embedded atom method ({RF-MEAM})
  energy and force-fitting code.
\newblock {\em Comput Phys Commun}, 196:439 -- 445, 2015.

\bibitem{Barrett2016}
C.~D. Barrett and R.~L. Carino.
\newblock The {MEAM} parameter calibration tool: an explicit methodology for
  hierarchical bridging between ab initio and atomistic scales.
\newblock {\em Integrat Mater Manuf Innov}, 5(1):9, 2016.

\bibitem{Stukowski2017}
A.~Stukowski, E.~Fransson, M.~Mock, and P.~Erhart.
\newblock Atomicrex—a general purpose tool for the construction of atomic
  interaction models.
\newblock {\em Model Simul Mater Sc}, 25(5):055003, 2017.

\bibitem{Horsfield-96}
A.~Horsfield, A.~M. Bratkovsky, M.~Fearn, D.~G. Pettifor, and M.~Aoki.
\newblock Bond-order potentials: Theory and implementation.
\newblock {\em Phys Rev B}, 53:12694, 1996.

\bibitem{Hammerschmidt2019}
T.~Hammerschmidt, B.~Seiser, M.E. Ford, A.N. Ladines, S.~Schreiber, N.~Wang,
  J.~Jenke, Y.~Lysogorskiy, C.~Teijeiro, M.~Mrovec, M.~Cak, E.R. Margine, D.G.
  Pettifor, and R.~Drautz.
\newblock {BOPfox} program for tight-binding and analytic bond-order potential
  calculations.
\newblock {\em Comput Phys Commun}, 235:221 -- 233, 2019.

\bibitem{Plimpton-95}
S.~Plimpton.
\newblock Fast parallel algorithms for short-range molecular dynamics.
\newblock {\em J Comput Phys}, 117:1, 1995.

\bibitem{Larsen2017}
A.~Hjorth Larsen, J.~J. Mortensen, J.~Blomqvist, I.~E. Castelli,
  R.~Christensen, M.~Dułak, J.~Friis, M.~N. Groves, B.~Hammer, C.~Hargus,
  E.~D. Hermes, P.~C. Jennings, P.~B. Jensen, J.~Kermode, J.~R. Kitchin, E.~L.
  Kolsbjerg, J.~Kubal, K.~Kaasbjerg, S.~Lysgaard, J.~B. Maronsson, T.~Maxson,
  T.~Olsen, L.~Pastewka, A.~Peterson, C.~Rostgaard, J.~Schiøtz, O.~Schütt,
  M.~Strange, K.~S. Thygesen, T.~Vegge, L.~Vilhelmsen, M.~Walter, Z.~Zeng, and
  K.~W. Jacobsen.
\newblock The atomic simulation environment—a python library for working with
  atoms.
\newblock {\em J Phys Condens Mat}, 29(27):273002, 2017.

\bibitem{Teijeiro-16-2}
C.~Teijeiro, T.~Hammerschmidt, R.~Drautz, and G.~Sutmann.
\newblock Efficient parallelisation of analytic bond-order potentials for large
  atomistic simulations.
\newblock {\em Comp Phys Comm.}, 204:64, 2016.

\bibitem{Teijeiro-16-1}
C.~Teijeiro, T.~Hammerschmidt, B.~Seiser, R.~Drautz, and G.~Sutmann.
\newblock Complexity analysis of simulations with analytic bond-order
  potentials.
\newblock {\em Model Simul Mater Sc}, 24:025008, 2016.

\bibitem{Jenke2019}
J.~Jenke, A.~N. Ladines, T.~Hammerschmidt, D.~G. Pettifor, and R.~Drautz.
\newblock Tight-binding bond parameters across the periodic table from
  downfolding the density-functional theory wave function.
\newblock {\em in preparation}, 2019.

\bibitem{Krishnapriyan2017}
A.~Krishnapriyan, P.~Yang, A.~M.~N. Niklasson, and M.~J. Cawkwell.
\newblock Numerical optimization of density functional tight binding models:
  Application to molecules containing carbon, hydrogen, nitrogen, and oxygen.
\newblock {\em J Chem Theory Comput}, 13(12):6191--6200, 2017.

\bibitem{Scipy}
E.~Jones, T.~Oliphant, P.~Peterson, et~al.
\newblock {SciPy}: Open source scientific tools for {Python}, 2001--.

\bibitem{Nlopt}
S.~Johnson.
\newblock The {NLopt} nonlinear-optimization package, 2001--.

\bibitem{Madsen2011}
G.~K.~H. Madsen, E.~J. McEniry, and R.~Drautz.
\newblock Optimized orthogonal tight-binding basis: Application to iron.
\newblock {\em Phys Rev B}, 83:184119, 2011.

\bibitem{Kresse-96-1}
G.~Kresse and J.~Furthm{\"uller}.
\newblock Efficiency of ab-initio total energy calculation for metals and
  semiconductors using a plane-wave basis set.
\newblock {\em Comp Mater Sci}, 6:15, 1996.

\bibitem{Kresse-96-2}
G.~Kresse and G.~Furthm{\"u}ller.
\newblock Efficient iterative schemes for ab initio total-energy calculations
  using a plane-wave basis set.
\newblock {\em Phys Rev B}, 54:11169, 1996.

\bibitem{Hammerschmidt-13}
T.~Hammerschmidt, A.~F. Bialon, D.~G. Pettifor, and R.~Drautz.
\newblock Topologically close-packed phases in binary transition-metal
  compounds: matching high-throughput ab-initio calculations to an empirical
  structure-map.
\newblock {\em New J Phys}, 15:115016, 2013.

\bibitem{Perdew-96}
J.P. Perdew, K.~Burke, and M.~Ernzerhof.
\newblock Generalized gradient approximation made simple.
\newblock {\em Phys Rev Lett}, 77:3865, 1996.

\bibitem{Bloechl-94}
P.~Bl\"{o}chl.
\newblock Projector augmented-wave method.
\newblock {\em Phys Rev B}, 50:17953, 1994.

\bibitem{Monkhorst-76}
H.~J. Monkhorst and J.~D. Pack.
\newblock Special points for {Brillouin}-zone integrations.
\newblock {\em Phys Rev B}, 13:5188, 1976.

\bibitem{Ladines2015}
A.N. Ladines, T.~Hammerschmidt, and R.~Drautz.
\newblock Structural stability of {Fe}-based topologically close-packed phases.
\newblock {\em Intermetallics}, 59:59 -- 67, 2015.

\bibitem{Golesorkhtabar2013}
R.~Golesorkhtabar, P.~Pavone, J.~Spitaler, P.~Puschnig, and C.~Draxl.
\newblock Elastic: A tool for calculating second-order elastic constants from
  first principles.
\newblock {\em Comput Phys Commun}, 184(8):1861 -- 1873, 2013.

\bibitem{Togo2015}
A.~Togo and I.~Tanaka.
\newblock First principles phonon calculations in materials science.
\newblock {\em Scripta Mater}, 108:1--5, Nov 2015.

\bibitem{Paidar1999}
V.~Paidar, L.~G. Wang, M.~Sob, and V.~Vitek.
\newblock A study of the applicability of many-body central force potentials in
  {NiAl} and {TiAl}.
\newblock {\em Model Simul Mater Sc}, 7(3):369--381, 1999.

\bibitem{Rayne1961}
J.~A. Rayne and B.~S. Chandrasekhar.
\newblock Elastic constants of iron from 4.2 to 300$^\circ$k.
\newblock {\em Phys Rev}, 122:1714--1716, 1961.

\bibitem{DeSchepper1983}
L.~De~Schepper, D.~Segers, L.~Dorikens-Vanpraet, M.~Dorikens, G.~Knuyt, L.~M.
  Stals, and P.~Moser.
\newblock Positron annihilation on pure and carbon-doped
  $\ensuremath{\alpha}$-iron in thermal equilibrium.
\newblock {\em Phys Rev B}, 27:5257--5269, 1983.

\bibitem{Vehanen1982}
A.~Vehanen, P.~Hautoj\"arvi, J.~Johansson, J.~Yli-Kauppila, and P.~Moser.
\newblock Vacancies and carbon impurities in $\ensuremath{\alpha}$- iron:
  Electron irradiation.
\newblock {\em Phys Rev B}, 25:762--780, 1982.

\bibitem{Domain2001}
C.~Domain and C.~S. Becquart.
\newblock Ab initio calculations of defects in fe and dilute fe-cu alloys.
\newblock {\em Phys Rev B}, 65:024103, 2001.

\bibitem{Blonski2007}
P.~Błoński and A.~Kiejna.
\newblock Structural, electronic, and magnetic properties of bcc iron surfaces.
\newblock {\em Surf Sci}, 601(1):123 -- 133, 2007.

\bibitem{Klotz2000}
S.~Klotz and M.~Braden.
\newblock Phonon dispersion of bcc iron to 10 gpa.
\newblock {\em Phys Rev Lett}, 85:3209--3212, 2000.

\bibitem{Jenke2018}
J.~Jenke, A.~P.~A. Subramanyam, M.~Densow, T.~Hammerschmidt, D.~G. Pettifor,
  and R.~Drautz.
\newblock Electronic structure based descriptor for characterizing local atomic
  environments.
\newblock {\em Phys Rev B}, 98:144102, 2018.

\end{thebibliography}

\end{document}